\documentclass[letterpaper, 10pt, conference, twocolumn]{IEEEtran}
\usepackage{fancyhdr}
\usepackage{amsmath,epsfig}
\usepackage{threeparttable}
\usepackage{epsf,epsfig}
\usepackage{amsmath}
\usepackage{amssymb}
\usepackage{amsfonts}
\usepackage[noadjust]{cite}
\usepackage{dsfont}
\usepackage{subfigure}

\begin{document}

\newtheorem{theorem}{\it Theorem}
\newtheorem{acknowledgement}[theorem]{Acknowledgement}
\newtheorem{axiom}[theorem]{Axiom}
\newtheorem{case}[theorem]{Case}
\newtheorem{claim}[theorem]{Claim}
\newtheorem{conclusion}[theorem]{Conclusion}
\newtheorem{condition}[theorem]{Condition}
\newtheorem{conjecture}[theorem]{Conjecture}
\newtheorem{criterion}[theorem]{Criterion}
\newtheorem{definition}[theorem]{Definition}
\newtheorem{example}[theorem]{Example}
\newtheorem{exercise}[theorem]{Exercise}
\newtheorem{lemma}{Lemma}
\newtheorem{corollary}{Corollary}
\newtheorem{notation}[theorem]{Notation}
\newtheorem{problem}[theorem]{Problem}
\newtheorem{proposition}{Proposition}
\newtheorem{solution}[theorem]{Solution}
\newtheorem{summary}[theorem]{Summary}
\newtheorem{assumption}{Assumption}
\newtheorem{examp}{\bf Example}
\newtheorem{probform}{\bf Problem}
\def\remark{{\noindent \bf Remark:\hspace{0.5em}}}

\def\qed{$\Box$}
\def\QED{\mbox{\phantom{m}}\nolinebreak\hfill$\,\Box$}
\def\proof{\noindent{\emph{Proof:} }}
\def\poof{\noindent{\emph{Sketch of Proof:} }}
\def
\endproof{\hspace*{\fill}~\qed
\par
\endtrivlist\unskip}
\def\endproof{\hspace*{\fill}~\qed\par\endtrivlist\vskip3pt}

\def\E{\mathbf{E}}
\def\eps{\varepsilon}
\def\phi{\varphi}
\def\Lsp{{\boldsymbol L}}
\def\Bsp{{\boldsymbol B}}
\def\lsp{{\boldsymbol\ell}}
\def\Ltsp{{\Lsp^2}}
\def\Lpsp{{\Lsp^p}}
\def\Linsp{{\Lsp^{\infty}}}
\def\LtR{{\Lsp^2(\Rst)}}
\def\ltZ{{\lsp^2(\Zst)}}
\def\ltsp{{\lsp^2}}
\def\ltZt{{\lsp^2(\Zst^{2})}}
\def\ninN{{n{\in}\Nst}}
\def\oh{{\frac{1}{2}}}
\def\grass{{\cal G}}
\def\ord{{\cal O}}
\def\dist{{d_G}}
\def\conj#1{{\overline#1}}
\def\ntoinf{{n \rightarrow \infty }}
\def\toinf{{\rightarrow \infty }}
\def\tozero{{\rightarrow 0 }}
\def\trace{{\operatorname{trace}}}
\def\ord{{\cal O}}
\def\UU{{\cal U}}
\def\rank{{\operatorname{rank}}}
\def\acos{{\operatorname{acos}}}

\def\SINR{\mathrm{SINR}}
\def\SNR{\mathrm{SNR}}
\def\SIR{\mathrm{SIR}}

\setcounter{page}{1}

\newcommand{\eref}[1]{(\ref{#1})}
\newcommand{\fig}[1]{Fig.\ \ref{#1}}

\def\bydef{:=}
\def\ba{{\mathbf{a}}}
\def\bb{{\mathbf{b}}}
\def\bc{{\mathbf{c}}}
\def\bd{{\mathbf{d}}}
\def\bee{{\mathbf{e}}}
\def\bff{{\mathbf{f}}}
\def\bg{{\mathbf{g}}}
\def\bh{{\mathbf{h}}}
\def\bi{{\mathbf{i}}}
\def\bj{{\mathbf{j}}}
\def\bk{{\mathbf{k}}}
\def\bl{{\mathbf{l}}}
\def\bm{{\mathbf{m}}}
\def\bn{{\mathbf{n}}}
\def\bo{{\mathbf{o}}}
\def\bp{{\mathbf{p}}}
\def\bq{{\mathbf{q}}}
\def\br{{\mathbf{r}}}
\def\bs{{\mathbf{s}}}
\def\bt{{\mathbf{t}}}
\def\bu{{\mathbf{u}}}
\def\bv{{\mathbf{v}}}
\def\bw{{\mathbf{w}}}
\def\bx{{\mathbf{x}}}
\def\by{{\mathbf{y}}}
\def\bz{{\mathbf{z}}}
\def\b0{{\mathbf{0}}}

\def\bA{{\mathbf{A}}}
\def\bB{{\mathbf{B}}}
\def\bC{{\mathbf{C}}}
\def\bD{{\mathbf{D}}}
\def\bE{{\mathbf{E}}}
\def\bF{{\mathbf{F}}}
\def\bG{{\mathbf{G}}}
\def\bH{{\mathbf{H}}}
\def\bI{{\mathbf{I}}}
\def\bJ{{\mathbf{J}}}
\def\bK{{\mathbf{K}}}
\def\bL{{\mathbf{L}}}
\def\bM{{\mathbf{M}}}
\def\bN{{\mathbf{N}}}
\def\bO{{\mathbf{O}}}
\def\bP{{\mathbf{P}}}
\def\bQ{{\mathbf{Q}}}
\def\bR{{\mathbf{R}}}
\def\bS{{\mathbf{S}}}
\def\bT{{\mathbf{T}}}
\def\bU{{\mathbf{U}}}
\def\bV{{\mathbf{V}}}
\def\bW{{\mathbf{W}}}
\def\bX{{\mathbf{X}}}
\def\bY{{\mathbf{Y}}}
\def\bZ{{\mathbf{Z}}}

\def\mA{{\mathbb{A}}}
\def\mB{{\mathbb{B}}}
\def\mC{{\mathbb{C}}}
\def\mD{{\mathbb{D}}}
\def\mE{{\mathbb{E}}}
\def\mF{{\mathbb{F}}}
\def\mG{{\mathbb{G}}}
\def\mH{{\mathbb{H}}}
\def\mI{{\mathbb{I}}}
\def\mJ{{\mathbb{J}}}
\def\mK{{\mathbb{K}}}
\def\mL{{\mathbb{L}}}
\def\mM{{\mathbb{M}}}
\def\mN{{\mathbb{N}}}
\def\mO{{\mathbb{O}}}
\def\mP{{\mathbb{P}}}
\def\mQ{{\mathbb{Q}}}
\def\mR{{\mathbb{R}}}
\def\mS{{\mathbb{S}}}
\def\mT{{\mathbb{T}}}
\def\mU{{\mathbb{U}}}
\def\mV{{\mathbb{V}}}
\def\mW{{\mathbb{W}}}
\def\mX{{\mathbb{X}}}
\def\mY{{\mathbb{Y}}}
\def\mZ{{\mathbb{Z}}}

\def\cA{\mathcal{A}}
\def\cB{\mathcal{B}}
\def\cC{\mathcal{C}}
\def\cD{\mathcal{D}}
\def\cE{\mathcal{E}}
\def\cF{\mathcal{F}}
\def\cG{\mathcal{G}}
\def\cH{\mathcal{H}}
\def\cI{\mathcal{I}}
\def\cJ{\mathcal{J}}
\def\cK{\mathcal{K}}
\def\cL{\mathcal{L}}
\def\cM{\mathcal{M}}
\def\cN{\mathcal{N}}
\def\cO{\mathcal{O}}
\def\cP{\mathcal{P}}
\def\cQ{\mathcal{Q}}
\def\cR{\mathcal{R}}
\def\cS{\mathcal{S}}
\def\cT{\mathcal{T}}
\def\cU{\mathcal{U}}
\def\cV{\mathcal{V}}
\def\cW{\mathcal{W}}
\def\cX{\mathcal{X}}
\def\cY{\mathcal{Y}}
\def\cZ{\mathcal{Z}}
\def\cd{\mathcal{d}}
\def\Mt{M_{t}}
\def\Mr{M_{r}}
\def\O{\Omega_{M_{t}}}
\newcommand{\figref}[1]{{Fig.}~\ref{#1}}
\newcommand{\tabref}[1]{{Table}~\ref{#1}}

\newcommand{\var}{\mathrm{Var}}
\newcommand{\fb}{\tx{fb}}
\newcommand{\nf}{\tx{nf}}
\newcommand{\BC}{\tx{(bc)}}
\newcommand{\MAC}{\tx{(mac)}}
\newcommand{\Pout}{P_{\tx{out}}}
\newcommand{\nnn}{\nn\\}
\newcommand{\FB}{\tx{FB}}
\newcommand{\TX}{\tx{TX}}
\newcommand{\RX}{\tx{RX}}
\renewcommand{\mod}{\tx{mod}}
\newcommand{\m}[1]{\mathbf{#1}}
\newcommand{\td}[1]{\tilde{#1}}
\newcommand{\sbf}[1]{\scriptsize{\textbf{#1}}}
\newcommand{\stxt}[1]{\scriptsize{\textrm{#1}}}
\newcommand{\suml}[2]{\sum\limits_{#1}^{#2}}
\newcommand{\sumlk}{\sum\limits_{k=0}^{K-1}}
\newcommand{\eqhsp}{\hspace{10 pt}}
\newcommand{\tx}[1]{\texttt{#1}}
\newcommand{\Hz}{\ \tx{Hz}}
\newcommand{\sinc}{\tx{sinc}}
\newcommand{\tr}{\mathrm{tr}}
\newcommand{\diag}{\mathrm{diag}}
\newcommand{\MAI}{\tx{MAI}}
\newcommand{\ISI}{\tx{ISI}}
\newcommand{\IBI}{\tx{IBI}}
\newcommand{\CN}{\tx{CN}}
\newcommand{\CP}{\tx{CP}}
\newcommand{\ZP}{\tx{ZP}}
\newcommand{\ZF}{\tx{ZF}}
\newcommand{\SP}{\tx{SP}}
\newcommand{\MMSE}{\tx{MMSE}}
\newcommand{\MINF}{\tx{MINF}}
\newcommand{\RC}{\tx{MP}}
\newcommand{\MBER}{\tx{MBER}}
\newcommand{\MSNR}{\tx{MSNR}}
\newcommand{\MCAP}{\tx{MCAP}}
\newcommand{\vol}{\tx{vol}}
\newcommand{\ah}{\hat{g}}
\newcommand{\tg}{\tilde{g}}
\newcommand{\teta}{\tilde{\eta}}
\newcommand{\heta}{\hat{\eta}}
\newcommand{\uh}{\m{\hat{s}}}
\newcommand{\eh}{\m{\hat{\eta}}}
\newcommand{\hv}{\m{h}}
\newcommand{\hh}{\m{\hat{h}}}
\newcommand{\Po}{P_{\mathrm{out}}}
\newcommand{\Poh}{\hat{P}_{\mathrm{out}}}
\newcommand{\Ph}{\hat{\gamma}}
\newcommand{\mat}[1]{\begin{matrix}#1\end{matrix}}
\newcommand{\ud}{^{\dagger}}
\newcommand{\C}{\mathcal{C}}
\newcommand{\nn}{\nonumber}
\newcommand{\nInf}{U\rightarrow \infty}

\title{\huge \setlength{\baselineskip}{30pt} Overlaid Cellular and Mobile Ad Hoc Networks}
\author{\authorblockN{Kaibin Huang\authorrefmark{1}, Yan Chen\authorrefmark{2}\authorrefmark{1}, Bin Chen\authorrefmark{3}, Xia Yang\authorrefmark{3}, and Vincent K. N. Lau\authorrefmark{1}}
\authorblockA{\authorrefmark{1}
Department of Electronic \& Computer Engineering\\
Hong Kong University of Science \& Technology\\
 Clear Water Bay, Hong Kong\\
Email: khuang@ieee.org, eeknlau@ust.hk}
\authorblockA{
\authorrefmark{2}Institute of Information \& Communication Engineering\\
Zhejiang University, Hangzhou, 310027, P.R. China\\
Email: yanchen@ust.hk}
\authorblockA{
\authorrefmark{3}Huawei Technologies Co. Ltd.\\
Bantian, Longgang District, Shenzhen 518129, P.R.China\\
Email: binchen@huawei.com, yangxia@huawei.com}
\vspace{-25pt}}

\maketitle
\
\begin{abstract}
In cellular systems using frequency division duplex, growing Internet services cause unbalance of uplink and downlink traffic, resulting in poor uplink spectrum utilization. Addressing this issue, this paper considers overlaying an ad hoc network onto a cellular uplink network for improving spectrum utilization and spatial reuse efficiency. Transmission capacities of the overlaid networks are analyzed, which are defined as the maximum densities of the ad hoc nodes and  mobile users under an outage constraint. Using tools from stochastic geometry, the capacity tradeoff curves for the overlaid networks are shown to be linear. Deploying overlaid networks based on frequency separation is proved to achieve higher network capacities than that based on spatial separation. Furthermore, spatial diversity is shown to enhance network capacities.

\end{abstract}

\section{Introduction}\label{Section:Intro}
In existing cellular systems based on frequency division duplex (FDD) such as FDD UMTS \cite{UMTS}, equal bandwidths are usually allocated for uplink and downlink transmission. With rapidly growing wireless Internet services, downlink traffic load is increasingly heavier than the uplink counterpart, which is a key characteristic of Internet data \cite{Marques:OppUse3GUplink:2008, KimJeong:CapUnbalanceULDL:CDMA:2000}. Consequently, uplink spectrum is underutilized. To address this issue, this paper considers overlaying an ad hoc network onto a FDD cellular system for improving the utilization of uplink spectrum and also the spatial reuse efficiency. This paper focuses on characterizing the network capacity tradeoff between the overlaid cellular and ad hoc networks.


One measure of network capacity  is \emph{transport capacity} for multi-hop networks that was introduced in  \cite{GuptaKumar:CapWlssNetwk:2000} and  has attracted active research (see e.g. \cite{Jovicic:UppBndTranpCapWlssNetwk:2004, AyferTse:HierCoopOptimCapAdHocNetwk:2006}). Another measure of network capacity is \emph{transmission capacity} introduced in \cite{WeberAndrews:TransCapWlssAdHocNetwkOutage:2005} for single-hop networks, which is defined as the maximum number of successful communication links  per unit area under the signal-to-interference-noise ratio (SINR) and outage constraints. Transmission capacity has been used for investigating the spatial reuse efficiency of different types of mobile ad hoc networks (see e.g. \cite{WeberAndrews:TransCapAdHocNetwkDistSch:2006, Huang:SpatialInterfCancel:PerCSI:Globecom08}). Compared with transport capacity, transmission capacity allows more accurate analysis and easier computation \cite{WeberAndrews:TransCapWlssAdHocNetwkOutage:2005}, and hence is adopted in this paper.

This paper considers overlaying the cellular uplink and ad hoc networks using two methods. The first is \emph{blind transmission} where the transmission of ad hoc nodes and mobile users are independent; the second is \emph{frequency mutual exclusion} where ad hoc nodes transmit over frequency sub-channels unoccupied by mobile users. For both methods, the tradeoff curves for the transmission capacities of two overlaid networks are shown to be linear. The region bounded by each curve and positive axes is called the capacity region, which contains all allowable combinations of ad hoc node and mobile user densities under the outage constraint.  The capacity region for frequency mutual exclusion is proved to be larger than that for blind transmission. Thus, the former is a more efficient method for overlaying cellular and ad hoc networks. Furthermore, capacity regions for both overlay methods are shown to be enlarged by exploiting spatial diversity.

\section{Network and Channel Models}

\subsection{Network Model}\label{Section:NetworkModel}
\begin{figure}
\centering
  \includegraphics[width=8cm]{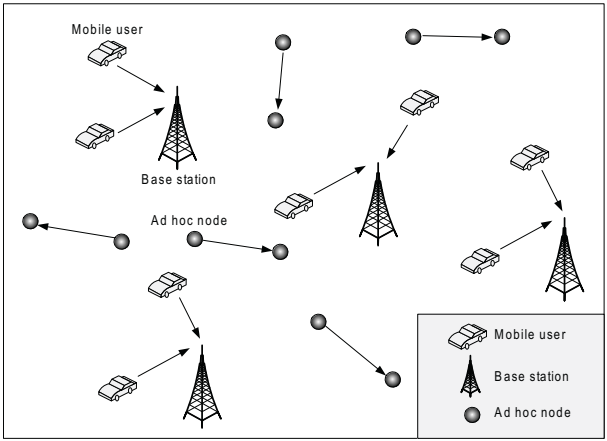}\vspace{10pt}\\
  \caption{Overlaid cellular and ad hoc networks}\label{Fig:Network}
\end{figure}

The overlaid cellular and ad hoc networks is illustrated in Fig.~\ref{Fig:Network}. The locations of transmitting nodes in the mobile ad hoc network are modeled as a Poisson point process (PPP) following the common approach in the literature \cite{WeberAndrews:TransCapWlssAdHocNetwkOutage:2005, WeberAndrews:TransCapWlssAdHocNetwkSIC:2005}. Specifically, the positions of the transmitters form a homogeneous PPP, denoted as $\Phi$,  on a $2$-dimensional plane and with the density represented by $\lambda_a$. Let $T_n$ denote the coordinate of the $n$th transmitting node. Each transmitting node is associated with a receiving node located at a fixed distance denoted as $d$.

Similarly, the positions of base stations in the cellular network are modeled as a homogeneous PPP on the same plane as the ad hoc network. Let $\Pi_b$ and $\lambda_b$ represent the process of base stations and its density, respectively. Each base station communicates with $M$ uplink mobile users, hereafter referred to simply as mobile users. Thus, the density of mobile users is $\lambda_u=M\lambda_b$. The coordinates of the $n$th base station and the $m$th associated mobile user are represented by $B_n$ and $U_{n,m} = B_n + Z_{n, m}$ respectively. To simplify analysis, the distance between a base station and a mobile user is fixed at $r$, thus $\left|Z_{n,m}\right| = r \ \forall \ m, n$. The case of random distances is considered in the full-length paper.

The uplink spectrum is divided into $K$ sub-channels. Each mobile user occupies one sub-channel. To simplify analysis, different base stations assign the same subset of $M$ sub-channels to their correspond sets of $M$ mobile users. Let $\mathcal{M}$ denote the indices of the $M$ sub-channels used by mobile users. Each transmitter in the ad hoc network either randomly selects a sub-channel for transmission from the complete set of sub-channels or the subset not being used by mobile users. These two methods are referred to as \emph{blind transmission} and \emph{frequency mutual exclusion}, respectively.

The networks are assumed to be interference limited and thus noise is neglected for simplicity. Consequently, the reliability of data packets received by either a base station or a node is determined by the signal-to-interference ratio (SIR). Let $S$ denote the random channel power for an arbitrary link, $\Omega$ the process of interferers\footnote{As discussed later, depending on the overlay method , the interferers can include unintended mobile users,  ad hoc transmitters in $\Phi$, or both.}, and $I_n$ the $n$th interferer in $\Omega$. Thus, assuming that data transmission power, denoted as $P_D$, is identical for all transmitters,  the SIR at the receiver end of the link under consideration is given as
\begin{equation}\label{Eq:SIR}
\SIR = \frac{S}{\sum_{T_n\in\Psi}I_n}.
\end{equation}
Since the $\SIR$ is independent of $P_D$, $P_D=1$ is assumed for simplicity. The correct decoding of received data packets requires the SIR to exceed a threshold $\theta$, which is identical for all receivers in the networks. In other words, the rate of information sent from a transmitter to a receiver is no less than $\log_2(1+\theta)$ assuming Gaussian signaling. To support this information rate with high probability, the outage probability that $\SIR$ is below $\theta$ must be smaller than or equal to a given threshold $0<\epsilon<1$, i.e.
\begin{equation}\label{Eq:Pout:Def}
\Pout(\lambda) = \Pr(\SIR < \theta) \leq \epsilon
\end{equation}
where $\Pout(\lambda)$ denotes the SIR outage probability as a function of $\lambda$.

\subsection{Channel Model}\label{Section:ChannelModel}
The model for a single sub-channel is described as follows, where the sub-channel index is omitted. Assume a narrow-band sub-channel model with flat fading. Each sub-channel consists of path-loss and small fading components. Consider a typical receiver which can be either a base station   or a ad hoc node.  A sub-channel between the typical receiver and the associated transmitter is $r^{-\alpha} Q_0$ or $d^{-\alpha} W_0$ depending on whether the transmitter is a mobile user or a node, where $\alpha$ and ($Q_0$ and $W_0$) represent the path-loss exponent and fading gains, respectively. Let $\Omega$ be the point process grouping all interferers to the typical receiver, including either mobile users, or ad hoc transmitters, or both depending on the network overlay method. Then a sub-channel between the $n$th interferer in $\Omega$ and the typical receiver is $D_n^{-\alpha}G_n$, where $D_n$ is the distance and $G_n$ is the fading gain. The random variables $\{D_n\}$ and $\{G_n\}$ are i.i.d. within the same set.

\section{Network Capacity Tradeoff: Asymptotic Analysis}
Similar to the notion of \emph{transmission capacity} introduced in \cite{WeberAndrews:TransCapWlssAdHocNetwkOutage:2005, WeberAndrews:TransCapAdHocNetwkDistSch:2006}, the network capacity considered in this paper is defined as the maximum density  of transmitters supported by a network under the outage constraint in \eqref{Eq:Pout:Def}. Let $P_b(\lambda_u, \lambda_a)$ and $P_a(\lambda_u, \lambda_a)$ denote the outage probabilities at the base station and an ad hoc receiver, respectively. Then the capacities of the cellular ($\lambda_u$) and the ad hoc ($\lambda_a$) networks can be written as
\begin{eqnarray}\label{Eq:Cap:Def}
(\lambda_u, \lambda_a) = \min\left(\arg P_b^{-1}(\epsilon), \arg P_a^{-1}(\epsilon)\right).
\end{eqnarray}
Note that the throughput per unit area for the cellular and the ad hoc networks are $(1-\epsilon)\log(1+\theta)\lambda_u$ and $(1-\epsilon)\log(1+\theta)\lambda_a$, respectively. The network capacity tradeoff between the overlaid cellular and ad hoc networks is analyzed for the regime of small outage probability regime $\epsilon \rightarrow 0$ in subsequent subsections.

\begin{figure}
\centering
  \includegraphics[width=8cm]{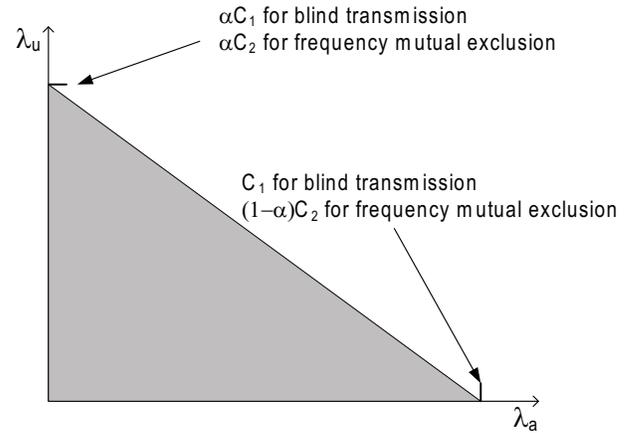}\\
  \caption{The network capacity region for the overlaid cellular ($\lambda_u$) nd ad hoc  ($\lambda_a$) networks in the regime of small outage probability. The parameter $\alpha$ is the percentage of sub-channels used by the cellular networks; $C_1$ and $C_2$ are defined in \eqref{Eq:CapReg:a} and \eqref{Eq:CapReg:b}, respectively.} \label{Fig:CapRegion}
\end{figure}

\subsection{Existing Results}
The capacity of a mobile ad hoc network under an outage constraint is analyzed in \cite{WeberAndrews:TransCapAdHocNetwkDistSch:2006}. The key results in \cite{WeberAndrews:TransCapAdHocNetwkDistSch:2006} are summarized in the following lemma and serve as the starting point for our analysis.

\begin{lemma}[Weber \emph{et al} \cite{WeberAndrews:TransCapAdHocNetwkDistSch:2006}]\label{Lem:Outage} Consider a narrow-band ad hoc network where the locations of transmitters are a homogeneous PPP having the density $\lambda$. The channel between a typical pair of transmitter and receiver is $d^{-\alpha}W$, and that between the typical receiver and the $n$th interferer is $D_n^{-\alpha}G_n$.
The SIR outage probability  for the typical receiver is bounded as
$\Pout^l \leq \Pout \leq \Pout^u$ where
\begin{eqnarray}
\Pout^l &=& 1- \E\left[\exp\left(- \kappa \lambda \theta^{\delta}\right)\right]\nn\\
\Pout^u &=& 1-\E\left[\left(1-\frac{\frac{\delta}{2-\delta}\kappa \lambda \theta^{-\delta}}{\left(1-\frac{\delta}{1-\delta}\kappa \lambda \theta^{-\delta}\right)^2}\right)^+e^{-\kappa \lambda \theta^{-\delta}}\right]\nn
\end{eqnarray}
where $\delta = \frac{2}{\alpha}$ and $\kappa = \pi\E[G^\delta] W^{-\delta} d^2$. For $\epsilon \rightarrow 0$, $\Pout$ can be expanded as\footnote{This result is generalized from that in \cite{WeberAndrews:TransCapAdHocNetwkDistSch:2006} where the asymptotic case $\theta \rightarrow \infty$ is considered. Note that $\epsilon\rightarrow 0$ corresponds to $\lambda \theta^{\delta}\rightarrow 0$. }
\begin{equation}\label{Eq:Pout:Asym}
\Pout = \E[\kappa] \lambda \theta^{\delta} + O\left(\lambda^2\theta^{2\delta}\right).
\end{equation}
\end{lemma}

\subsection{Capacity Tradeoff: Blind Transmission}
The outage probability for data received at a base station is derived for $\epsilon \rightarrow 0$. Consider the $m$th one of the $M$ sub-channels assigned to mobile users. The mobile users using this sub-channel form a point process denoted as $\Delta_m=\{B + Z_m\mid B\in \Phi\}=\Phi + Z_m$, where $Z_m$ represents the shift in location. Since the homogeneous PPP $\Phi$ is translation invariant, $\Delta_m$ is also a homogeneous PPP with the density $\lambda_b$. Consider a typical base station $B$ and the associated mobile user $U_m$ using the $m$th assigned sub-channel, where the subscript $n$ is omitted for simplicity. Without loss of generality, $B$ is assumed to be located at the origin, and hence $U_m = Z_m$. The data received at $B$ from $U_m$ is interfered with by other-cell mobile users and the transmitters in the ad hoc networks that  use the same sub-channel, which are represented by the processes $\{\Delta_m/\{U_m\}\}$ and $\Pi_m$, respectively. The process of these interferers is $\{\Delta_m/\{U_m\}\}\cup\Pi_m$. By Slivnyak's Theorem \cite{StoyanBook:StochasticGeometry:95}, conditioned on $U_m$, the process of other mobiles users $\{\Delta_m/\{U_m\}\}$ remains as a homogeneous PPP with the same density $\lambda_b$.

Next, consider a typical pair of transmitter $T$ and receiver $R$ in the ad hoc network. By blind transmission, $T$ and $R$ randomly selects one of $K$  sub-channel for communication. Let $S_T$ denote the index of the selected subchannel. Then $S_T$ follows the uniform distribution with the support $\{1, 2,\cdots, K\}$. By treating $S_T$ as a \emph{mark} of $T$ and using the Marking Theorem \cite{Kingman93:PoissonProc}, the marked PPP $\Pi_k = \{(T, S_T=k)\mid T\in \Pi\}$ is a homogeneous PPP with density $\lambda_a/K$. Combining above results and the superposition property of PPP, the process of interferers for $U_m$, namely  $\{\Delta_m/\{U_m\}\}\cup\Pi_m$, is a homogeneous PPP with the density $\lambda_b + \lambda_a$.

Based on the above results, the relationship $\lambda_u = \lambda_a/M$, and  Lemma~\ref{Lem:Outage}, for a single-user data stream received at a base station, the asymptotically small outage probability can be written as
\begin{equation}\label{Eq:Pout:Cellular}
P_b = \E[\kappa_b] \left(\frac{\lambda_u}{M} + \frac{\lambda_a}{K}\right)\theta^{\delta} + O\left(\left(\frac{\lambda_u}{M} + \frac{\lambda_a}{K}\right)^2\theta^{2\delta}\right).
\end{equation}
where $\kappa_b = \pi\E[G^\delta] Q^{-\delta} r^2$.

For a receiver in the ad hoc network, the outage probability is derived for $\epsilon \rightarrow 0$ as follows. Consider a typical pair of transmitter $T$ and receiver $R$ communicating using the $k$th sub-channel. If $k\notin\mathcal{M}$, the interferers for $R$ are only unintended transmitters in the ad hoc network that also use the $k$th sub-channel, represented by $\Pi_k/\{T\}$. If $k\in\mathcal{M}$, $R$ is interfered with by both the transmitters in $\Pi_k/\{T\}$ and also the mobile users using the $k$th sub-channel grouped in the process $\Delta_m$. Similar to earlier derivation, by using Lemma~\ref{Lem:Outage},  for the data received at a receiver in the ad hoc network using the $k$ sub-channel, the asymptotically small outage probability can be written as
\begin{equation}\label{Eq:Pout:AdHoc}
P_a(k) = \left\{\begin{aligned}
&\E[\kappa_a] \left(\frac{\lambda_u}{M} + \frac{\lambda_a}{K}\right) \theta^{\delta} + O\left(\left(\frac{\lambda_u}{M} + \frac{\lambda_a}{K}\right) ^2\theta^{2\delta}\right),\\
& k \in \mathcal{M}\\
&\E[\kappa_a] \left(\frac{\lambda_a}{K}\right) \theta^{\delta} + O\left(\left(\frac{\lambda_a}{K}\right)^2\theta^{2\delta}\right),\quad k \notin \mathcal{M}
\end{aligned}\right.
\end{equation}
where $\kappa_a = \pi\E[G^\delta] W^{-\delta} d^2$.

The tradeoff between the capacity of the cellular and ad hoc networks is obtained by using \eqref{Eq:Cap:Def}, \eqref{Eq:Pout:Cellular} and \eqref{Eq:Pout:AdHoc}. Denote asymptotic equivalence as $\cong$ and define it as the relationship between two functions $f_1(\epsilon)$ and $f_2(\epsilon)$ that satisfies $\lim_{\epsilon \rightarrow 0}\frac{f_1(\epsilon)}{f_2(\epsilon)}=1$. Recall that the capacities of the cellular and the ad hoc networks are defined in \eqref{Eq:Cap:Def} as the maximum densities of mobile users $\lambda_u$ and ad hoc transmitters $\lambda_a$ under the outage constraint. The network capacity tradeoff curve thus obtained is given in the following proposition.
\begin{proposition}\label{Prop:CapReg:Blind}
Using the overlay method of blind transmission,  the capacity of the cellular network $\lambda_u$ and that of the ad hoc network $\lambda_a$ satisfies the following relationship
\begin{equation}\label{Eq:CapReg:a}
\frac{\lambda_u}{\alpha} + \lambda_a \cong C_1, \quad \epsilon \rightarrow 0
\end{equation}
where $\alpha$ is the fraction of sub-channels assigned to mobile users, and
\begin{equation}\label{Eq:C1}
C_1 = \frac{K\epsilon \theta^{-\delta}}{\pi\E\left[G^\delta\right]}\min\left(\frac{1}{d^2\E[W^{-\delta}]}, \frac{1}{r^2\E[Q^{-\delta}]}\right).
\end{equation}
\end{proposition}
The region bounded by the curve in \eqref{Eq:CapReg:a} and the constraints $\lambda_a\geq 0$ and $\lambda_u\geq0$ is the capacity region for the overlaid cellular and ad hoc networks as illustrated in Fig.~\ref{Fig:CapRegion}. Any combination of $\lambda_a$ and $\lambda_u$ within this region satisfies the outage constraint in \eqref{Eq:Pout:Def}. As observed from Fig.~\ref{Fig:CapRegion}(a), the area of capacity region is increased by increasing $\alpha$. This implies that for blind transmission, allocating a subset of sub-channels rather than the whole set to mobile users results in inefficient use of the spectrum. Besides increasing the number of  mobile users, the capacity region can be also enlarged
by increasing $C_1$ in \eqref{Eq:C1}. This is equivalent to optimizing a set of parameters, including increasing the number of sub-channels $K$ or the outage constraint $\epsilon$, or reducing the distances between the transmitter and receiver in each link ($d$ and $r$) or the functions of fading gains ($\E[W^{-\delta}]$ and $\E[Q^{-\delta}]$). Reducing these functions can be achieved by exploiting spatial diversity as discussed in Section~\ref{Section:SpaDiv}.

\subsection{Capacity Tradeoff: Frequency Mutual Exclusion}
For frequency mutual exclusion, the cellular and ad hoc networks use different sub-sets of sub-channels, namely $\mathcal{M}$ and $\mathcal{K}/\mathcal{M}$, and thus do not interfere with each other. Specifically, the data stream received at a base station over the sub-channel $m\in\mathcal{M}$ contains interference from mobile users in other cells using the same sub-channels; a node receiving data using $k$th sub-channel is interfered with only by unintended transmitting nodes also using the sub-channel $k$. Following a similar procedure as in the preceding section, the outage probabilities for the overlaid networks are obtained as
\begin{eqnarray}
P_b &\!\!\!=\!\!\!& \frac{\E[\kappa_b] \lambda_u \theta^{\delta}}{M} + O\left(\lambda_u^2\theta^{2\delta}\right)\label{Eq:Pout:Cell:b}\\
P_a &\!\!\!=\!\!\!& \E[\kappa_a] \left(\frac{\lambda_a}{K-M}\right) \theta^{\delta} + O\left(\left(\frac{\lambda_a}{K-M}\right)^2\theta^{2\delta}\right)\label{Eq:Pout:AdHoc:b}
\end{eqnarray}
where $\kappa_a = \pi\E[G^\delta] W^{-\delta} d^2$.
By applying the outage constraint to \eqref{Eq:Pout:Cell:b} and \eqref{Eq:Pout:AdHoc:b}, the tradeoff curve for the capacities of the cellular and ad hoc networks is derived and given in the following proposition.
\begin{proposition}\label{Prop:CapReg:Excl}
Using the overlay method of frequency mutual exclusion,  the capacity of the cellular network $\lambda_u$ and that of the ad hoc network $\lambda_a$ satisfies the following relationship
\begin{equation}
\frac{\lambda_u}{\alpha} + \frac{\lambda_a}{1-\alpha} \cong C_2, \quad \epsilon \rightarrow 0 \label{Eq:CapReg:b}
\end{equation}
where
\begin{equation}
C_2 = \frac{K\epsilon \theta^{-\delta}}{\pi\E\left[G^\delta\right]}\left(\frac{1}{d^2\E[W^{-\delta}]}+\frac{1}{r^2\E[Q^{-\delta}]}\right).
\end{equation}
\end{proposition}
As shown in Fig.~\ref{Fig:CapRegion}, the capacity region for the case of frequency mutual exclusion is bounded by the curve defined in \eqref{Eq:CapReg:b} and the first quadrant of the coordinate system. Similar comments as for the case of blind transmission are also applicable for the present case. By comparing \eqref{Eq:CapReg:a} and \eqref{Eq:CapReg:b}, it can be observed that $C_2>C_1$ and the factor $(1-\alpha)$ increases the capacity $\lambda_a$. In other words, the capacity region for frequency mutual exclusion is larger than that for blind transmission. A comparison of capacity regions is given in Section~\ref{Section:Numerical}.

\subsection{Effect of Spatial Diversity}\label{Section:SpaDiv}
The capacity region of the overlaid cellular and ad hoc networks can be enlarged by enhancing the channel gains of data links $W$ and $Q$ by exploiting spatial diversity. Spatial diversity is created using multiple antennas and provides an effective way of counteracting fading. Among many available spatial diversity techniques, we consider the simplest one: beamforming \cite{PaulrajBook}. Assuming i.i.d. Rayleigh fading and beamforming at either the receiver or transmitter (one-sided), the channel gains $W$ and $Q$ follow the chi-squared distribution with the number of complex degrees of freedom equal to $L_1$ and $L_2$, respectively, which are referred to as diversity orders. The diversity order for the interference channel $G$ is assumed to be one. Based on the above model, the network capacity regions in Proposition~\ref{Prop:CapReg:Blind} and \ref{Prop:CapReg:Excl} have close-form expressions as given in the following lemma.

\begin{figure}
\centering
  \includegraphics[width=9cm]{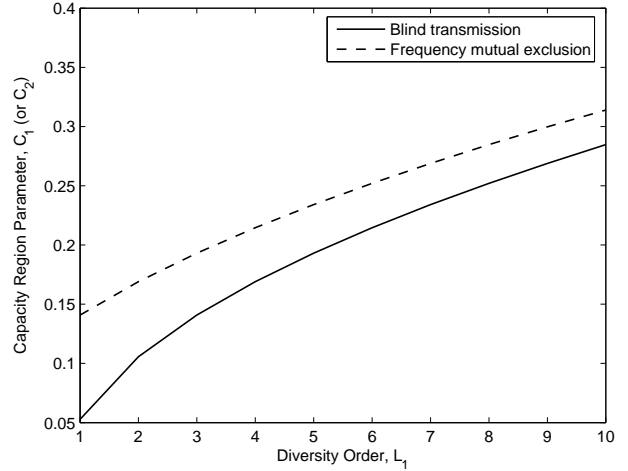}\\
  \caption{The capacity region parameter $C_1$ (or $C_2$) vs. diversity order $L_1$ for the data link in the ad hoc network. The data link in the cellular network has the diversity order $L_2=L_1+2$. Other parameters are chosen as $d=5$ meters, $\theta=3$, and $K=1000$ and $\epsilon = 0.01$.}\label{Fig:DivGain}
\end{figure}

\begin{lemma}\label{Lem:SpaDiv} For i.i.d. Rayleigh fading, data links using one-sided beamforming, and $r=d$, the capacity tradeoff curve is given as:
\begin{enumerate}
\item Blind transmission:
\begin{equation}\label{Eq:CapReg:c}
\frac{\lambda_u}{\alpha} + \lambda_a = \frac{\tilde{C}\Gamma(L_{\min})}{ \Gamma(L_{\min}-\delta)}, \quad \epsilon \rightarrow 0
\end{equation}
where $\tilde{C} = K\epsilon \theta^{-\delta}[\pi d^2 \Gamma(1+\delta)]^{-1}$.

\item Frequency mutual exclusion:
\begin{equation}\label{Eq:CapReg:d}
\frac{\lambda_u}{\alpha} + \frac{\lambda_a}{1-\alpha} = \tilde{C}\left[\frac{\Gamma(L_1)}{ \Gamma(L_1-\delta)}+\frac{\Gamma(L_2)}{ \Gamma(L_2-\delta)}\right], \quad \epsilon \rightarrow 0.\nn
\end{equation}

\end{enumerate}
\end{lemma}
The proof is given in Appendix. Let $C_1$ and $C_2$ denote the terms on the right-hand-size of \eqref{Eq:CapReg:c} and \eqref{Eq:CapReg:d}, respectively. For illustration, consider the diversity orders $L_2 = L_1+2$, the distance $r=d=5$ meters, the SIR threshold $\theta=3$, the number of sub-channels $K=1000$, and the outage constraint $\epsilon = 0.01$. Fig~\ref{Fig:DivGain} plots $C_1$ and $C_2$ versus the diversity order $L_1$. As observed from Fig~\ref{Fig:DivGain}, both $C_1$ and $C_2$ increases as $L_1$ grows due to the diversity gains. Moreover, $C_2$ is larger than $C_1$ and their difference is relatively larger for a smaller diversity order, indicating that the capacity for frequency mutual exclusion is higher than that for blind transmission.

\section{Simulation and Numerical Results}\label{Section:Numerical}

\begin{figure}
\centering
  \includegraphics[width=9cm]{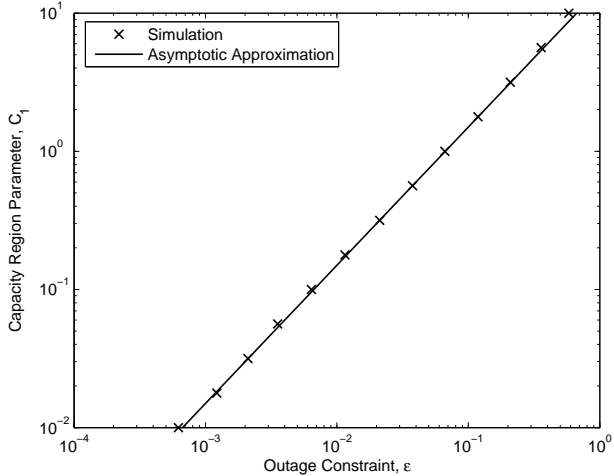}\\
  \caption{Comparison of the approximated and simulated values of the parameter $C_1$ that defines the capacity region for the overlaid cellular and ad hoc networks using blind transmission. The approximated values for $C_1$ are computed using \eqref{Eq:CapReg:a}.} \label{Fig:EvalAsymp}
\end{figure}

Fig.~\ref{Fig:EvalAsymp} compares the values of the capacity region parameter $C_1$ computed using \eqref{Eq:CapReg:a} and those obtained by simulation. The diversity order is chosen as $L_1 = L_2 = 4$ and other parameters have identical values as used for Fig.~\ref{Fig:DivGain}. As observed from Fig.~\ref{Fig:EvalAsymp}, the theoretical values obtained based on the asymptotical assumption ($\epsilon \rightarrow 0$) match the simulation results very well. A similar observation is also made for $C_2$ given in \eqref{Eq:CapReg:b}, but the detailed comparison is omitted in this paper. These observations lead to the conclusion that the asymptotical results obtain in Proposition~\ref{Prop:CapReg:Blind} and \ref{Prop:CapReg:Excl} are also applicable for the non-asymptotic regime of $\epsilon$ e.g. $\epsilon > 0.1$.

\begin{figure}
\centering
  \includegraphics[width=9cm]{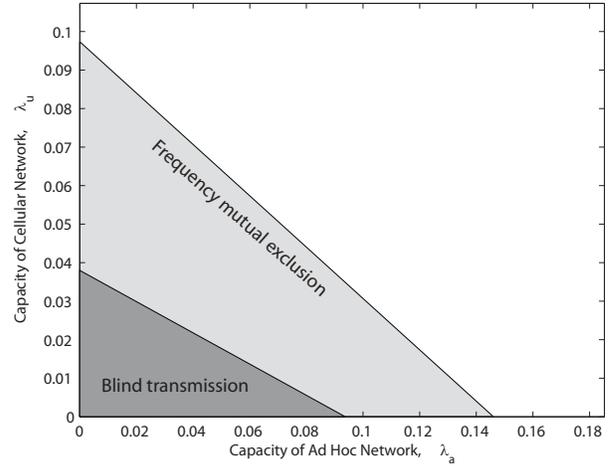}\\
  \caption{Comparison between the capacity regions for two network overlay methods, namely blind transmission and frequency mutual exclusion.}\label{Fig:CmpCapRegion}
\end{figure}

Fig.~\ref{Fig:CmpCapRegion} compares the capacity regions for two network overlay methods: blind transmission and frequency mutual exclusion. The values of parameters are identical to those used for generating Fig.~\ref{Fig:EvalAsymp}. As observed from Fig.~\ref{Fig:CmpCapRegion}, the network overlay based on frequency mutual exclusion results in a much larger capacity region than that using blind transmission. Therefore, the former is a more efficient method for implementing the overlaid cellular and ad hoc networks. Nevertheless, the method of frequency mutual exclusion requires nodes in the ad hoc network to detect sub-channels unused by mobile users, but such an operation is unnecessary for blind transmission. Detection of available sub-channels can be based on carrier sending or side information broadcast by base stations.

\appendix[Proof for Lemma~\ref{Lem:SpaDiv}]
The result for the case of frequency mutual exclusion is obtained by substituting $\E[G^{\delta}] = \Gamma(1+\delta)$, $\E[Q^{-\delta}]=\Gamma(L_2-\delta)$, and $\E[W^{-\delta}]=\Gamma(L_2-\delta)$ into \eqref{Eq:CapReg:b}.

To obtain the result for the case of blind transmission, it is sufficient to use \eqref{Eq:CapReg:a} and prove that $\frac{\Gamma(L_1)}{ \Gamma(L_1-\delta)} > \frac{\Gamma(L_2)}{ \Gamma(L_2-\delta)}$ if $L_1 > L_2$ and vise versa. Without loss of generality, assume $L_1>L_2$. Using the property  $\Gamma(1+x) = x\Gamma(x)$ and $\Gamma(n) = (n-1)!$
\begin{eqnarray}
\frac{\Gamma(L_1)}{ \Gamma(L_1-\delta)}\!\!\!\! &\!\!\!\!=\!\!\!\!& \!\!\!\! \frac{(L_1-1)(L_1-2)\cdots L_2}{(L_1-1-\delta)(L_1-2-\delta)\cdots (L_2-\delta)}\frac{\Gamma(L_2)}{\Gamma(L_2-\delta)}\nn\\
&\overset{(a)}{>}& \frac{\Gamma(L_2)}{\Gamma(L_2-\delta)}.\nn
\end{eqnarray}
The inequality (a) holds since $\delta >0$. This completes the proof.

\bibliographystyle{ieeetr}

\end{document}